\documentclass[twocolumn,preprintnumbers,amsmath,amssymb, superscriptaddress]{revtex4}

\usepackage{graphicx}
\usepackage{dcolumn}
\usepackage{bm}
\usepackage{bbm}
\usepackage{amsmath}
\usepackage{amsthm}
\usepackage{dsfont}
\usepackage{amssymb}

 \renewenvironment{quote}{%
  \list{}{%
    \leftmargin1.5em   
    \rightmargin1.5em 
  }
  \item\relax
}
{\endlist}

\newcommand{\bra}[1]{\langle #1|}
\newcommand{\ket}[1]{|#1\rangle}

\newcommand{\ketbra}[2]{|#1\rangle\!\langle#2|}
\newcommand{\id}{\mathbbm{1}}

\let\oldmarginpar\marginpar
\renewcommand\marginpar[1]{\-\oldmarginpar[\raggedleft\marginparsize #1]%
{\raggedright\marginparsize #1}}

\begin{document}

\setlength{\tabcolsep}{1ex}

\title{Particle exchange in post-quantum theories}
\author{Oscar C.O. Dahlsten}
\affiliation{Atomic and Laser Physics, Clarendon Laboratory,
University of Oxford, Parks Road, Oxford OX1~3PU, United Kingdom}
\affiliation{Center for Quantum Technologies, National University of Singapore, Republic of Singapore}

\author{Andrew J.P. Garner}
\affiliation{Atomic and Laser Physics, Clarendon Laboratory,
University of Oxford, Parks Road, Oxford OX1~3PU, United Kingdom}

\author{Jayne Thompson}
\affiliation{Center for Quantum Technologies, National University of Singapore, Republic of Singapore}

\author{Mile Gu}
\affiliation{Center for Quantum Information, Interdisciplinary Institute of Information Sciences, Tsinghua University, Beijing, China}
\affiliation{Center for Quantum Technologies, National University of Singapore, Republic of Singapore}

\author{Vlatko Vedral}
\affiliation{Atomic and Laser Physics, Clarendon Laboratory,
University of Oxford, Parks Road, Oxford OX1~3PU, United Kingdom}
\affiliation{Center for Quantum Technologies, National University of Singapore, Republic of Singapore}

\date{\today}

\begin{abstract}
In quantum theory, particles in three spatial dimensions come in two different types: bosons or fermions, which exhibit sharply contrasting behaviours due to their different exchange statistics.
Could more general forms of probabilistic theories admit more exotic types of particles?
Here, we propose a thought experiment to identify more exotic particles in general post-quantum theories.
We consider how in quantum theory the phase introduced by swapping indistinguishable particles can be measured.
We generalise this to post-quantum scenarios whilst imposing indistinguishability and locality principles.
We show that our ability to witness exotic particle exchange statistics depends on which symmetries are admitted within a theory.
These exotic particles can manifest unusual behaviour, such as non-abelianicity even in topologically simple three-dimensional space.
\end{abstract}



\maketitle

{\bf Introduction.---}
In quantum theory, particles in three spatial dimensions come in two different types: bosons or fermions.
What, however, if quantum theory did not always hold?
There could be broader theories, of which quantum theory is a subset, or an unencountered physical regime governed by a different theory altogether. In such a theory, what kind of particles could we have?
Our interest in this is two-fold: firstly to understand better why quantum particles come in the types they do,  and secondly, to search for potential post-quantum physical phenomena.

We tackle this question by identifying how exotic particle statistics may be detected operationally.
We propose an operational protocol where the particles in question are physically swapped conditioned on the state of some reference system.
The swap induces a relative phase with respect to the reference that has observational consequence.
Our treatment does not assume quantum theory.
In particular, we adopt the {\em convex framework}~\cite{Hardy01,BarnumBLW06,Barrett07,Mana03,MasanesM11}.
This framework models physical reality purely through detector statistics, and has already demonstrated much success in studying potential non-local correlations beyond quantum theory~\cite{Tsirelson93, PopescuR94, Barrett07}.

We identify the observational effects of swapping two indistinguishable particles conditioned on an ancillary system in this general framework.
From the foundational principles of locality and indistinguishability we derive natural constraints on how such systems can transform.
In particular, the effect of swapping indistinguishable particles can be gauged by measurements which compare the reference system before and after the swap.
This comparison can be facilitated by swapping the indistinguishable particles conditioned on a binary measurement of an ancillary system; here locality arguments prohibit allowed transformations from changing the statistics of the binary measurement.
We say the allowed transformations belong to the {\em phase group} of the binary measurement,
the group of reversible operations on the system which do not alter the statistics associated with this measurement for any state~\cite{GarnerDNMV13}.
What remains is the possibility that there is {\em one set of particle exchange statistics for each element of a phase group of a binary measurement in the theory}.

One may associate some of the phase group elements with generalised bosons (corresponding to the trivial phase group identity element), and others to generalised fermions and even anyons (which exhibit arbitrary statistics under exchange~\cite{LeinaasM77, Wilczek82, NayakSSFS08, Pachos12}).
Beyond quantum theory there can be several types of fermionic statistics, and moreover in theories with a non-abelian phase group, these fermions might manifest non-commutative behaviour under swapping.
This is conceptually different from non-abelian anyons arising from braiding considerations in lower dimensional quantum scenarios (see e.g.~\cite{NayakSSFS08}); even when considering the exchange of these generalised fermions in path-independent scenarios (particularly in topologically simple theories where swapping twice has no measurable effect) they can still demonstrate non-abelian behaviour.


{\em \bf Framework and tools.---}
To go beyond the quantum case, we consider the framework of convex probabilistic theories, sometimes known as the {\em generalised probabilistic theory} (GPT) framework~\cite{Hardy01,Mana03,BarnumBLW06,Barrett07,MasanesM11}.

The convex framework amounts to taking the minimalistic pragmatic view that the operational content of a theory is contained in the predicted statistics of measurement outcomes. Essentially any experiment generating a data table can be represented in this framework\cite{Hardy01,Mana03}.

As in quantum (and classical probability) theory, the central role is played by the {\em states} of a system, along with the measurement outcomes.
The states and the measurement outcomes are represented by vectors of real numbers, such that the probability of measuring a particular outcome on a particular state is given by the inner product of these vectors.

As an illustration, we mention how the quantum case can be treated in this way.
There are several schemes for representing density matrices (and other Hermitian operators) as real vectors, which do fit the framework, e.g.\ the Bloch sphere representation of a qubit: one may write $\rho=\sum_{i}\xi_i \sigma_i$ such that $\xi_i\in\mathbbm{R}\,\forall i$ and $(\sigma_0, \sigma_1, \sigma_2, \sigma_3)=(\id, X, Y, Z)$, referring to the identity and the standard Pauli matrices.
The state may then be represented as the real vector $(\xi_0,\xi_1,\xi_2, \xi_3)$ or a vector proportional to this.
Similarly, projection operators are represented by the same such real vectors as would represent the corresponding pure state $\ket{\psi}\bra{\psi}$.
The probability of a particular outcome for a given measurement in the standard quantum representation is $\mathrm{Tr}(\rho \Pi)$, which coincides with the Euclidean inner product of the real vectors.

The set of allowed states is known as the {\em state space}.
These sets are convex, as the name of the framework would suggest.
In quantum physics, the state space of a qubit is the Bloch sphere.
For non-quantum theories, which might not be describable by an underlying mathematical structure such as a Hilbert space, the state space might take on an arbitrary convex shape which could admit vectors outside the Bloch sphere, corresponding to states forbidden in quantum theory (such as the state where it is possible to perfectly predict the outcome of all three of the $X$, $Y$ and $Z$ measurements).

A transformation, $T$ on states is given by a matrix of real numbers (see e.g.~\cite{Hardy01}).
A necessary criterion for $T$ to be an allowed transformation is that it takes all allowed states to allowed states.
$T$ is said to be {\em reversible} if both $T$ and its matrix inverse, $T^{-1}$, are allowed transformations.
The set of allowed reversible transformations in a theory is contained within (and might be) the automorphism group of the theory's state space.

A {\em theory} in the framework is specified by the sets of allowed measurement-outcome vectors, allowed states and allowed transformations.

We shall also use the notion of the {\em phase group} in convex probabilistic theories~\cite{GarnerDNMV13}. A phase group is a subgroup of the group of reversible transformations, defined by demanding that the elements preserve the statistics of a chosen measurement for all states.


{\em \bf Quantum version of swap experiment.---}
Consider two particles in a pure state $\ket{AB}$.
Our argument makes no distinction between fundamental or composite particles, nor between particles or quasi-particles.
The particles are  indistinguishable if physically swapping them (e.g.~having two separate traps with one particle in each and moving them into each other's previous position) does not change anything observable about the state.
Denoting the physical swap operation by the transformation $\pi_{A\leftrightarrow B}$ we demand $\pi_{A\leftrightarrow B}\ket{AB}=e^{i\theta}\ket{AB}$.
This picture of active particle swapping inducing phase change is consistent with the example of anyonic particle statistics arising by taking an anyon to be a ring of charges with a magnetic flux through it, such that the phase induced by actively swapping two anyons arises from the Aharonov-Bohm effect~\cite{Wilczek82, Pachos12}.

\begin{figure}[tb]
\centering
\includegraphics[width=0.9\linewidth]{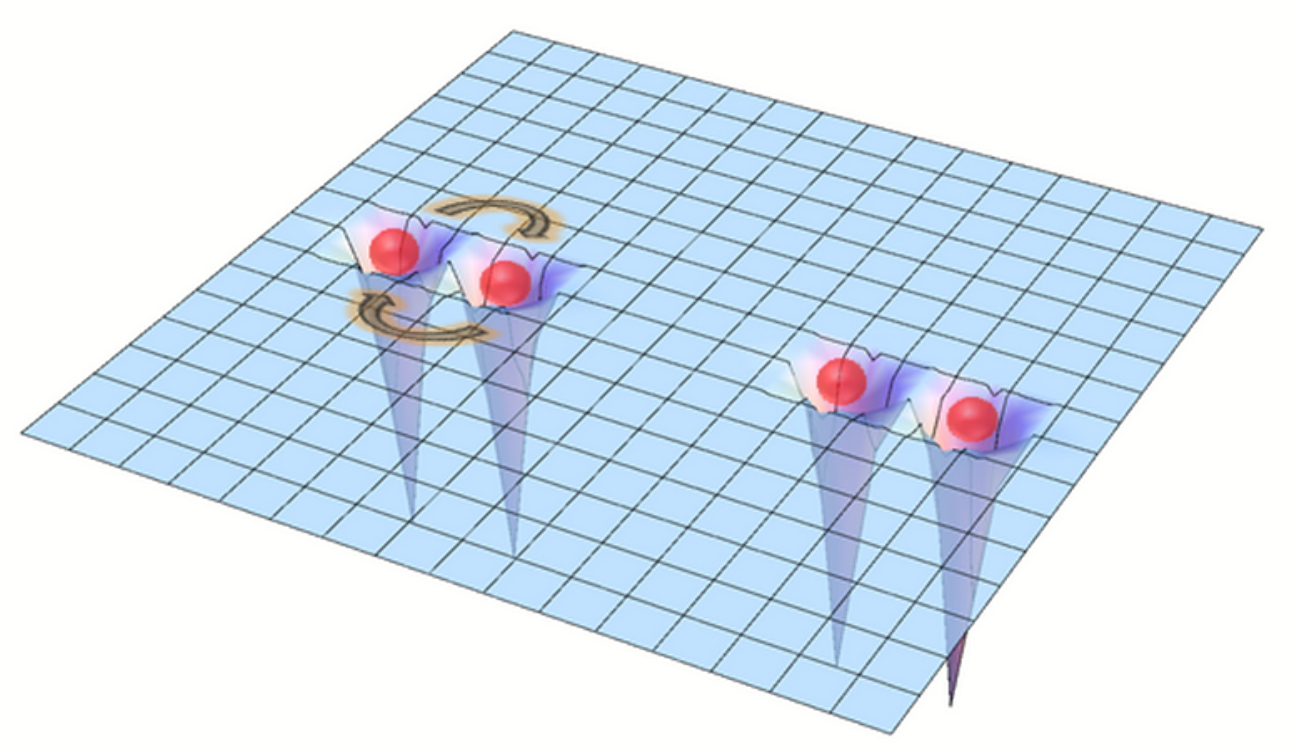}
\caption{{\em A physical scenario for particle exchange}. The two particles are trapped such that there is at most one particle in each well with negligible probability of any particle being found in between traps or jumping from one to the other. The wells on the left are space-like separated relative to those on the right. The total state is in a superposition of two scenarios: (i) there is one particle in each of the wells on the left, (ii) there is one particle in each of the wells on the right. On the left side the wells are swapped in an active, physical, transformation. In the quantum case this would introduce a relative phase between (i) and (ii) which could be determined experimentally. We consider what could happen if the systems were indistinguishable post-quantum particles. }
\label{fig:interferometer}
\end{figure}

Whilst the phase introduced by the swap is global and not observable, it does have observable affects by introducing an ancillary reference system, $C$. We may then do the physical swap {\em controlled} by the third system, meaning that in the branch of the superposition where $C$ is in some state $\ket{0}$ no swap is done and no phase introduced, but in the branch of the superposition where $C$ is in $\ket{1}$ the swap is done and a phase introduced. Mathematically, the (unnormalised) state evolves as:
\begin{equation*}
(\ket{0}+\ket{1})\ket{AB}\rightarrow \ket{0}\ket{AB}+e^{i\theta}\ket{1}\ket{AB}=(\ket{0}+e^{i\theta}\ket{1})\ket{AB}.
\end{equation*}
One may think of this as being realized through the idealised experiment in figure~\ref{fig:interferometer}.

This results in a phase transformation on the reference system $C$, $U(\theta) = \ketbra{0}{0}+e^{i\theta}\ketbra{1}{1}$.
Thus, in general we may associate a phase transformation of a qubit with each quantum particle type.

In principle there may be several topologically distinct ways of swapping two particles which could all give rise to different phases.
These topological constraints can impose a condition on the behaviour after repeated swaps, which generates a sub-group from the phase group.
A particularly important condition is to impose that that performing the swap twice amounts to the identity operation, such that $e^{i\theta}=\pm 1$, generating the sub-group containing elements corresponding to the boson fermion cases respectively.
This restriction follows from the fact that swapping particles twice is equivalent to orbiting one about the other, and in theories with three or more spatial dimensions, this is topologically equivalent to doing nothing at all~\cite{NayakSSFS08}.
In two dimensions this is not always possible, enabling the existence of anyons.

{\em \bf Post-quantum version of swap experiment.---}
We now consider the same set-up, described using the convex framework outlined above.
We shall say that particles have different types when the experiment yields distinguishable outcome statistics for the respective inputs.

The state of the system is now represented by a real vector (in the special case of quantum theory representing a density matrix). We wish to swap two indistinguishable particles $A$ and $B$.
Let $\varphi_{AB}$ represent the combined operational state of particles $A$ and $B$.
Crucially, we posit that swapping $A \leftrightarrow B$ {\em leaves the state $\varphi_{AB}$ invariant}, whether the two particles form a closed system or are part of a larger system.
We take this condition as the {\em definition} of what it means for $A$ and $B$ to be indistinguishable.
To observe the effect of a swap we introduce a third system $C$. We construct a combined state for the three systems: $\varphi_{ABC}$ and define a new operation $T$ which corresponds to swapping the particles $A$ and $B$.

To examine the possible particle statistics observable in the experiment we take the input states to be products
\begin{equation}\label{eq:inputstate}
\varphi_{ABC}^{in} = \varphi_{AB}^{in} \otimes \varphi_C^{in}
\end{equation}
where $\varphi_{AB}$ and $\varphi_C$ are pure states.
The operational meaning of this product is that the probabilities of measurements on either side factorise (see appendix).
Such product states are convenient for our analysis and should exist in any reasonable description of joint systems.

For indistinguishable particles the output state
\begin{eqnarray}
\varphi_{ABC}^{out} &=&  \varphi_{AB}^{out} \otimes \varphi_C^{out},\nonumber \\
&=&  \varphi_{AB}^{in} \otimes \varphi_C^{out}.
\end{eqnarray}
The first line follows from noting that by the assumption of reversibility $\varphi_{AB}^{out}$ is pure and that Theorem 2 of~\cite{Hardy09} shows that such states are uncorrelatable. The second line follows from the definition of indistinguishability which implies that the marginal state on AB is invariant under the swap.

The above argument implies that the effect of the swap must be entirely reflected in the final state of $\varphi_C^{out}$ (a generalisation of quantum ``phase kick-back'').
Now, with $A$ and $B$ indistinguishable, $T$ acts on the state (\ref{eq:inputstate}) as
\begin{equation}
T(\varphi_{ABC}^{in}) = \left({\id}_{AB} \otimes T_C \right) (\varphi_{ABC}^{in})
\end{equation}
for some $T_C$ acting on $C$ exclusively. (This does not imply that T acts as $T = {\id}_{AB} \otimes T_C$ on all states).

It is possible to take locality considerations into account within the GPT framework~\cite{DahlstenGV12}, and use this to argue why {\em $T_C$ must be in the phase group} associated with a binary measurement on C.
Recall that the physical procedure involves two space-like separated locations, as in figure~\ref{fig:interferometer}.
The swap is done on one location only. Let $Z$ be the binary observable distinguishing the two branches, left vs. right.
By locality, the swap cannot change the $Z$-statistics (the branches could be space-like separated, and super-luminal signalling would be possible if the probability of being in one branch could be changed by an action, or lack of, on the other). Thus the transformation must be in the phase group associated with the $Z$ measurement. We now arrive at our main claim:
\begin{quote}
{\em The observable effect of exchanging two indistinguishable particles is restricted to the control system being transformed by an element of the phase group of a binary measurement. Hence the particle types perceivable in a theory correspond to elements of this phase group.}
\end{quote}

{\bf Discussion.---}
A particularly surprising consequence of our reasoning is that non-abelian particle types might exist for some theories even in three-dimensional space with trivial topological properties.
We enter this trivial regime by imposing the restriction that swapping twice has no effect, which generates a sub-group containing only the identity (which we associate with {\em bosonic} behaviour) and the elements which are self inverse (which we associate with {\em fermionic} behaviour).
(All particle types outside this sub-group could be considered a generalisation of anyonic behaviour).
In general, theories with a phase group containing more than one self-inverse element have the potential to display non-abelian statistics.

\begin{figure}[tb]
\centering
\includegraphics[width=0.75\linewidth]{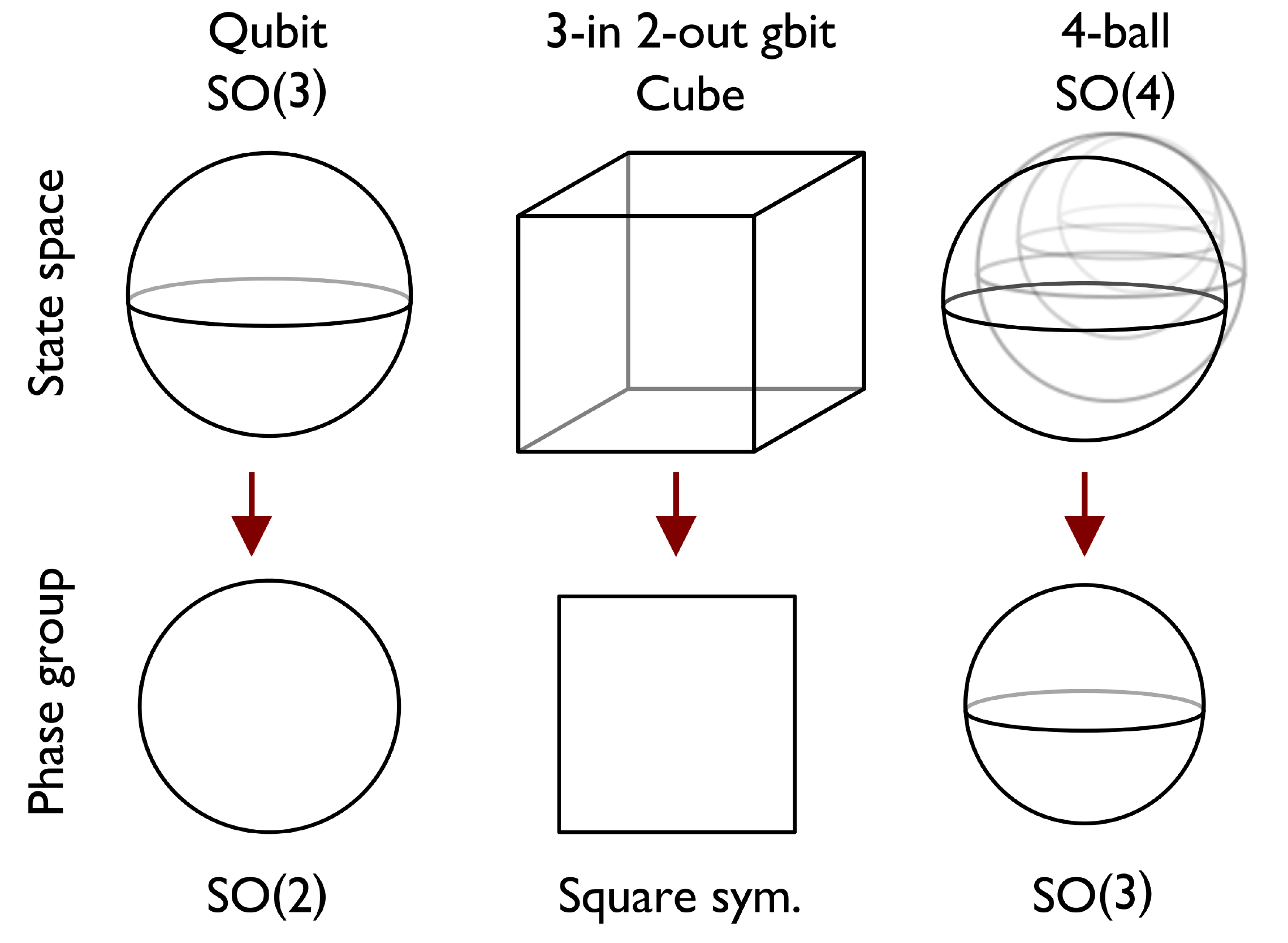}
\caption{
{\em Illustrated examples of phase groups associated with 2-level systems, as defined in~\cite{GarnerDNMV13}}. (A system is called two-level if 2 pure states and no more can be perfectly distinguished in a single measurement). Our argument associates one particle type with each element of such a phase group. In the post-quantum case these groups may be non-abelian by virtue of including rotations in several dimensions, as well as reflections. (To have a quantum non-abelian phase group, the measurement needs to be coarse grained, implying a less fundamental type of non-abelianicity.)
}
\label{fig:phasegroups}
\end{figure}

For example, consider a modified quantum theory with four independent measurements (see Figure~\ref{fig:phasegroups}).
Here, it is possible to rotate a spin in at least three possible orthogonal directions, without affecting its measurement statistics in the fourth.
The phase group resulting from such transformations is isomorphic to $SU(2)$, and thus is clearly non-abelian.
However, even after taking the subgroup of self inverse elements, we are left with the set $\{\id, \sigma_x,\sigma_y,\sigma_z\}$ (where $\sigma_i$ is a $2\times2$ Pauli matrix), and $[\sigma_i,\sigma_j]\neq0$ for $i,j=1,2,3$, $i\neq j$.
Each of these Pauli matrices may be associated with a different type of particle, and so in this theory there are {\bf non-abelian fermions}.

Such exotic non-abelian particles demonstrate remarkable behaviour.
Imagine swapping two pairs of exotic particles consecutively (for example, in the above theory, X-ons and Y-ons associated with the $\sigma_x$ and $\sigma_y$ elements respectively), with both operations conditioned on a single two level control system.
Should the phase transformations associated with swapping these pairs not commute, which set is swapped first will have an observable consequence on the final state of the control system.
However, in a scenario where these pairs are always swapped by reversible operations (without a control bit) the order of operations should have no effect on the final state, as these swaps act on different non-overlapping subspaces.
This seems paradoxical, given that physical intuition would dictate that if two operations commute, they should remain commutative when conditioned on a coinciding degree of freedom (this is always the case in quantum theory, see appendix).
This remarkable deviation from intuition may offer a clear phenomenological indicator for post-quantum behaviour.

This result also highlights how theories with more complex phase groups can admit a greater diversity of possible particle types.
A particularly interesting subtlety is that the different types of particle we can observe is dependent on the control system we use to `observe' them with.
This is not as counter-intuitive as it first sounds.
Consider attempting to distinguish between swapping two quantum bosons, or two quantum fermions, using only a classical control bit.
Such a bit does not allow any super-position states; the only classical dynamic allowed which preserves the statistics of the  only measurement allowed on the bit is to do nothing: the phase group consists only of the identity element~\cite{GarnerDNMV13}.
Thus, whether the particles swapped are actually bosons or actually fermions, we have no way of telling them apart by the measurement statistics on this classical bit.
Extending this argument, we note then that {\em even if more exotic particle types exist, if our experiment to detect them relies only on a quantum control bit, we can only detect quantum particle behaviour}.
This follows from the very definition of indistinguishablity being our ability to tell things apart.

The connection between phase group and particle types has dual applications.
On the one hand, it identifies a plethora of possible exotic particle types as we consider further generalizations of quantum theory (See Figure~\ref{fig:phasegroups} for a survey).
On the other, it forces us to confront the seemingly paradoxical properties of such particles should we concede that our reality is not completely described by quantum statistics.
Indeed, our connection motivates a natural method to quantify {\em how exotic} different theories are, based on their capacity to distinguish different types of exotic particles.
Classical theories, for example, would be considered the least exotic, as their phase group is trivial~\cite{GarnerDNMV13}.
Meanwhile, post-quantum theories could be far more exotic as not only would our observations depend on the types of particles that are exchanged, but also the order in which exchanges take place, even in three-dimensional space.
Ultimately, this could motivate arguments that such dependence on order is unphysical based on some underlying principle; and use this to provide an alternative argument justifying nature's preference for vanilla quantum mechanics.

{\em \bf Acknowledgements.---}We are grateful for discussions with Felix~Pollock in particular and also for discussions with Yuval~Gefen, Libby~Heaney, Markus~Johansson, Markus~M\"uller, Mio~Murao, Yoshifumi~Nakata and Jonathan~Oppenheim.
This research was funded by the National Research Foundation (Singapore), the Ministry of Education (Singapore), the EPSRC (UK), the Templeton Foundation, the Leverhulme Trust, the Oxford~Martin~School, Academic Research Fund Tier 3 MOE2012-T3-1-009, the National Basic Research Program of China Grant 2011CBA00300, 2011CBA00302 and the National Natural Science Foundation of China Grant 61033001, 61061130540.

\appendix

\bibliographystyle{h-physrev}
\bibliography{partstats_refs}

\begin{thebibliography}{10}

\bibitem{Hardy01}
L.~Hardy,
\newblock e-print  (2001), arXiv:quant-ph/0101012.

\bibitem{BarnumBLW06}
H.~Barnum, J.~Barrett, M.~Leifer, and A.~Wilce,
\newblock \prl {\bf 99}, 240501 (2007).

\bibitem{Barrett07}
J.~Barrett,
\newblock Phys. Rev. A {\bf 75}, 32304 (2007).

\bibitem{Mana03}
P.~{Mana},
\newblock e-print  (2003), 0305117v3.

\bibitem{MasanesM11}
L.~Masanes and M.~P. M{\"u}ller,
\newblock New Journal of Physics {\bf 13}, 063001 (2011).

\bibitem{Tsirelson93}
B.~S. Tsirelson,
\newblock Hadronic Journal Supplement {\bf 8}, 329 (1993).

\bibitem{PopescuR94}
S.~Popescu and D.~Rohrlich,
\newblock Found. Phys. {\bf 24}, 379 (1994).

\bibitem{GarnerDNMV13}
A.~J.~P. Garner, O.~C.~O. Dahlsten, Y.~Nakata, M.~Murao, and V.~Vedral,
\newblock New Journal of Physics {\bf 15}, 093044 (2013).

\bibitem{LeinaasM77}
J.~Leinaas and J.~Myrheim,
\newblock Il Nuovo Cimento B Series 11 {\bf 37}, 1 (1977).

\bibitem{Wilczek82}
F.~{Wilczek},
\newblock Phys. Rev. Lett. {\bf 49}, 957 (1982).

\bibitem{NayakSSFS08}
C.~Nayak, S.~H. Simon, A.~Stern, M.~Freedman, and S.~Das~Sarma,
\newblock Rev. Mod. Phys. {\bf 80}, 1083 (2008).

\bibitem{Pachos12}
J.~Pachos,
\newblock {\em {Introduction to Topological Quantum Computation}} (Cambridge
  University Press, Cambridge, 2012).

\bibitem{Hardy09}
L.~Hardy,
\newblock e-print  (2009), 0912.4740.

\bibitem{DahlstenGV12}
O.~C.~O. Dahlsten, A.~J.~P. Garner, and V.~Vedral,
\newblock e-print  (2012), arXiv:quant-ph/1206.5702.

\end{thebibliography}


\newpage
\enlargethispage{2\baselineskip}

\section*{Technical Appendix}
{\bf The convex framework.---}
Here we outline some properties of the convex framework so that our results might be more rigorously understood.

{\em States and effects.---}
For a given state $s$, the probability of a measurement outcome $e$ is calculated by combining the associated vectors $\vec{e}$ and $\vec{s}$ according to $p(e|s)=\vec{e}\cdot \vec{s}$, where the dot is the standard Euclidean inner product.
A crucial self-consistency condition that must hold in all theories is that for all allowed states $\vec{s}$ and measurement-outcome vectors $\vec{e}$, $0 \leq \vec{s}\cdot\vec{e}\leq 1$ so that the inner product {\em always} gives a probability.
A special effect, $\vec{1}$, measures the normalisation of a state, such that $\vec{1}\cdot\vec{s}=1$ if $\vec{s}$ is normalised.

{\em Convexity.---}
Probabilistic combinations of allowed states are also allowed states, corresponding to $\vec{s}=\sum_i p_i\vec{s_i}$, where state $s_i$ is prepared with probability $p_i$. This implies that the set of allowed states in a theory is always {\em convex}. As in quantum theory, states which are non-trivial probabilistic combinations of other states are termed {\em mixed} and the other states are termed {\em pure}.

{\em Linearity of transformations.---}
We demand that transformations on states respect such mixtures, such that if two states $s_1$, $s_2$ are prepared with probabilities $p_1$ and $p_2$, applying a transformation to each of them and then mixing the outcomes should be the same as applying the transformation to the mixture, $p_1 \vec{s_1}+p_2 \vec{s_2}$.
Mathematically, this demands $T(p_1 \vec{s_1}+p_2 \vec{s_2})=p_1 T\vec{s_1}+p_2 T\vec{s_2}$, which implies that $T$ is affine and can be represented by a matrix of real numbers~\cite{Hardy01}.

{\em Phase group.---}
In the quantum case, the simplest system that can be associated with a binary measurement is a qubit.
The set of unitary transformations of the form $U=e^{i\phi_0}\ket{0}\bra{0}+e^{i\phi_1}\ket{1}\bra{1}$ (corresponding to SO(2) azimuthal rotations around the Bloch sphere) do not alter the outcomes of the measurement $\{\ket{0}\bra{0},\ket{1}\bra{1}\}$, and so this set is in the {\em phase group} of this measurement.

A phase group in the convex framework more generally is the subgroup of the group of reversible transformations defined by demanding that the elements preserve the statistics of a chosen measurement for all states~\cite{GarnerDNMV13}.
Consider a given theory in the framework with a state space and group of allowed reversible transformations, $G$. Let $\{\vec{e}_i\}_{i=1}^M$ be a measurement, i.e. $\sum_i \vec{e}_i \cdot \vec{s}=1$ for all normalised states $\vec{s}$. The {\it phase group} $G_\Phi$ associated with the measurement is the maximal subgroup of all transformations $T\in G$ that leave all outcome probabilities of the measurement invariant, that is, for any states $\vec{s}$ and $\forall i$: $\vec{e_i} \cdot \vec{s}=\vec{e_i} \cdot (T \vec{s}).$ There are also irreversible phase dynamics important for decoherence~\cite{GarnerDNMV13}, but to avoid unnecessary technicality we shall restrict ourselves to reversible operations.

{\em An example post-quantum system.---}
One simple example post-quantum system is the {\em gbit} (generalised bit)~\cite{Barrett07}.
This arises as the system carrying the conditional marginal states of a Popescu-Rohrlich box (PR-box), a hypothetical system which has more Bell violation than allowed by quantum theory~\cite{PopescuR94,Tsirelson93,Barrett07}.
Each preparation of the gbit system is associated with a state $\vec{s} = \left(p(+1|X),\,p(-1|X),\,p(+1|Z),\,p(-1|Z)\right)$ where the measurement choices and outcomes are respectively labelled X/Z and $\pm 1$ in analogy with the quantum Pauli matrices.
(This scenario is called 2-in 2-out, meaning 2 measurements and 2 outcomes).
All such states can be written as probabilistic combinations of four states $\{(1,\, 0,\, 1,\, 0),(1,\, 0,\, 0,\, 1),(0,\, 1,\, 1,\, 0),(0,\, 1,\, 0,\, 1)\}$.
These four pure states form the corners of a square.
In the absence of any other constraints, the symmetries of the square may be taken as the allowed reversible transformations in such a theory~\cite{Barrett07}.

{\em Product states.---}
For two systems, $\varphi_A$ and $\varphi_B$, we can construct a joint state $\varphi_{AB}$, such that when we take any measurement $M$ on $\varphi_A$ with effects $\{\vec{m_i}\}$ and any measurement $N$ on $\varphi_B$ with effects $\{\vec{n_i}\}$, the probability $P(M\!=\!m_i, N\!=\!n_i | \varphi_{AB}) = P(M\!=\!m_i|\varphi_A) P(N\!=\!n_i|\varphi_B)$ for all $\varphi_A$ and $\varphi_B$.
Thus if we create a set of effects for our joint measurement $MN$ as $\vec{m_i}\otimes\vec{n_i}$ and the combined state $\varphi_A\otimes\varphi_B$, where $\otimes$ is the (e.g.) Kronecker product, then we note that the joint probability $P(M\!=\!m_i, N\!=\!n_i | \varphi_{A}\otimes\varphi_{B}) = (\vec{m_i}\otimes\vec{n_i})\cdot(\varphi_A\otimes\varphi_B) = (\vec{m_i}\cdot\varphi_A)\otimes(\vec{n_i}\cdot\varphi_B) = P(M\!=\!m_i|\varphi_A) P(N\!=\!n_i|\varphi_B)$. (The middle step follows by expressing the dot product $\vec{a}\cdot\vec{b}$ as a matrix multiplication $\vec{a}^\mathrm{T}\vec{b}$.)
Thus if we can express a state as such a tensor product, the joint probabilities can be factorised into probabilities on its sub-systems.

{\bf Controlled commuting quantum operators.---} 
In quantum theory if unitary operators $U$ and $V$ commute, controlled versions of these operators will also commute.
This is seen by noting that a simultaneously diagonalisable basis, $\{w_i\}$ exists for all such commuting unitary matrices, such that $U\ket{w_i}=u_i\ket{w_i}$ and $V\ket{w_i}=v_i\ket{w_i}$.
For the controlled versions of such operators $U_C$ ($V_C$), if the external control bit is $\ket{0}$ then $\id$ is applied to the system, and if it is $\ket{1}$ then $U$ ($V$) is applied.
The composite system is spanned by the basis $\{\ket{0},\ket{1}\}\otimes\{\ket{w_1}\ldots \ket{w_n}\}$, and so we can define the action of the two operators as $U_C: \ket{0}\otimes\ket{w_i} \to e^{i\phi_0} \ket{0}\otimes\ket{w_i} \forall i$, $\ket{1}\otimes\ket{w_i} \to e^{i\phi_1} \ket{1}\otimes u_i \ket{w_i} \forall i$ and likewise $V_C: \ket{0}\otimes\ket{w_i} \to e^{i\varphi_0} \ket{0}\otimes\ket{w_i} \forall i$, $\ket{1}\otimes\ket{w_i} \to e^{i\varphi_1} \ket{1}\otimes v_i \ket{w_i} \forall i$. ($\phi_0$, $\phi_1$, $\varphi_0$ and $\varphi_1$ are arbitrary phases between the control and target systems that may be introduced by a controlled operator whilst still allowing for the correct behaviour of the target subsystem.)
By inspection, we see that $U_C$ and $V_C$ are simultaneously diagonalisable in this basis too, and therefore must also commute.

{\bf Swapping with a classical control bit.---}
By embedding a classical bit in the quantum formalism, we can see that a classical control bit can not be used to distinguish between bosons and fermions.
Consider a classical control bit $\rho_c$ as a probabilistic mixture of $0$ or $1$: $\rho_c=P\ketbra{0}{0}+(1-P)\ketbra{1}{1}$.
As argued in the text, the only observable effect of a controlled swap operation is a phase transform on the control bit: if the particles are bosons, $U_b = \id$ is applied to $\rho_c$, and if they are fermions $U_f = \sigma_z$ is applied. However for classical $\rho_c$ because $U_f^\dag \rho_c U_f = U_b^\dag \rho_c U_b$ there is no way to tell which of these operations was applied by measuring the output. Hence classical bits can not distinguish between bosons and fermions.

\end{document}